\documentclass[doublecol,linenumbers]{epl2} % for 2 columns style with line numbers
\usepackage{multirow}
\usepackage{amssymb}
\usepackage{amsmath}
\usepackage{cancel}
\usepackage{verbatim}
\usepackage{subcaption}
\usepackage[utf8]{inputenc}

\title{Improved constraints on lepton asymmetry from the \\ cosmic microwave background}
\shorttitle{Improved CMB constraints on lepton asymmetry} %Insert here a short version of the title if it exceeds 70 characters

\author{Isabel M. Oldengott \inst{1} \and Dominik J. Schwarz\inst{1}}
\shortauthor{I. Oldengott and D. Schwarz}

\institute{                    
  \inst{1} Fakult\"at f\"ur Physik, Bielefeld University, D--33501 Bielefeld, Germany\\
}

\abstract{
A cosmic lepton asymmetry $\eta_{\text{l}}=(n_{\text{l}}-n_{\bar{\text{l}}})/n_{\gamma}$ affects the primordial helium 
abundance and the expansion rate of the early Universe. Both of these effects have an impact on the anisotropies of the 
cosmic microwave background (CMB). We derive constraints on the neutrino chemical potentials from the Planck 2015 data, 
assuming equal lepton flavour asymmetries and negligible neutrino masses. We find 
$\xi=-0.002 ^{+0.114}_{-0.111}$ (95\% CL) for the chemical potentials, which corresponds to 
$ -0.085 \leq \eta_{\text{l}} \leq 0.084$. Our constraints on the lepton asymmetry are significantly stronger than previous constraints from CMB data analysis and we argue that they are more robust than those from primordial light element abundances.  The resulting constraints on the primordial helium and deuterium abundances are consistent with those from direct measurements. 
}

\begin{document}

\maketitle

\section{Introduction}
\label{Introduction}
The lepton asymmetry of the Universe is one of the most weakly constrained cosmological parameters. Whereas the 
baryon asymmetry $\eta_{\text{b}} \equiv (n_{\text{b}}-n_{\bar{\text{b}}})/n_{\gamma} = 
274 \times 10^{-10} \omega_{\text{b}}= (6.099\pm 0.044) \times 10^{-10}$ \cite{Ade:2015xua} is tightly constrained, 
the lepton asymmetry \mbox{$\eta_{\text{l}} \equiv (n_{\text{l}}-n_{\bar{\text{l}}})/n_{\gamma}$} could be larger by 
many orders of magnitude. Here $n_i$ denote the number densities of particles of type $i$ and 
$\omega_{\text{b}} = \Omega_{\text{b}} h^2$, which is the dimensionless baryonic energy density $\Omega_{\text{b}}$ 
multiplied by the square of the dimensionless Hubble parameter $h$.

A large lepton asymmetry $\eta_{\text{l}}$ would be realised as an excess of the number density of neutrinos 
over anti-neutrinos or vice versa\footnote{In case of Majorana neutrinos, it would be an excess of right handed 
over left handed neutrinos or vice versa \cite{Langacker:1982fg}.} and is therefore possibly hidden in the cosmic 
neutrino background (C$\nu$B). For scenarios that invoke physics well above the electroweak scale to 
explain the matter-antimatter asymmetry of the Universe, typically 
a tiny lepton asymmetry, at the same order of magnitude as the baryon asymmetry, is favoured because 
sphaleron processes are expected to equilibrate the baryon and lepton asymmetry (e.g. \cite{Harvey:1990qw}). 
The existence of sphaleron processes is however not confirmed experimentally and numerous high energy 
theories predict a significantly larger lepton asymmetry, 
e.g.\ \cite{Linde:1976kh,Shi:1998km,MarchRussell:1999ig,McDonald:1999in,Eijima:2017anv}. 
Thus, lepton asymmetry is a key parameter to better constrain models of the creation of 
the matter-antimatter asymmetry of the Universe.

The tightest constraints on lepton asymmetry $|\eta_{\text{l}}| < \mathcal{O}(0.1)$ up to date are usually based on a combination of direct measurements of the primordial abundances of light elements with constraints on the baryon density from data of the cosmic microwave background (CMB), e.g. \cite{Simha:2008mt,Mangano:2011ip,Cooke:2013cba}. 

Independently from direct measurements of the primordial abundance of light elements, we can also derive constraints from the CMB alone, see e.g. \cite{Shiraishi:2009fu,Popa:2008tb,Schwarz:2012yw,Caramete:2013bua,Barenboim:2016lxv}. In this work, we derive updated constraints on lepton asymmetry using CMB data from the Planck satellite \cite{Ade:2015xua}. Our constraints are the tightest constraints on lepton asymmetry derived from CMB data up to date.

\section{Lepton asymmetry in the early Universe}

\begin{largetable}
\centering
\bgroup
\def\arraystretch{1.5}
\begin{tabular}{|c||c|c||c|}
\hline 
& & & (no lepton asymmetry) \\
 & TT+lowP+lensing & TTTEEE+lowP+lensing & TTTEEE+lowP+lensing \\ 
\hline 
\hline
$\xi$ & -0.011$ ^{+0.076}_{-0.098} \left( ^{+0.179}_{-0.167} \right)$ & -0.002 $ ^{+0.053}_{-0.060} \left( ^{+0.114}_{-0.111} \right)$ & 0 \\ 
\hline 
$\omega_{\text{b}}$ & 0.02235 $ ^{+0.00031}_{-0.00035} \left( ^{+0.00066}_{-0.00064} \right)$ & 0.02229 $ ^{+0.00022}_{-0.00023} \left( ^{+0.00043}_{-0.00044} \right) $ & 0.02228 $ ^{+0.00016}_{-0.00017} \left( ^{+0.00032}_{-0.00032} \right)$ \\ 
\hline 
$Y_{\text{p}}$ & 0.251 $ ^{+0.025}_{-0.020} \left( ^{+0.041}_{-0.042} \right) $ & 0.248  $ ^{+0.014}_{-0.014} \left( ^{+0.027}_{-0.027} \right) $ & 0.24784$^{+0.000067}_{-0.000069} \left(^{+0.000136}_{-0.000134} \right)$ \\ 
\hline 
$y_{\text{DP}}$ & 2.448 $^{+0.086}_{-0.084} \left( ^{+0.166}_{-0.166} \right)$& 2.443 $^{+0.055}_{-0.054} \left( ^{+0.107}_{-0.109} \right)$ & 2.439 $^{+0.029}_{-0.029} \left( ^{+0.057}_{-0.058} \right)$ \\
\hline
$n_{\text{s}}$ & 0.9704 $ ^{+0.011}_{-0.013} \left( ^{+0.024}_{-0.023} \right) $ & 0.9660 $ ^{+0.0078}_{-0.0083} \left( ^{+0.0158}_{-0.0159} \right) $ & 0.9654 $^{+0.0047}_{-0.0049} \left(^{+0.0096}_{-0.0094} \right)$  \\ 
\hline 
\end{tabular} 
\egroup
\caption{Cosmological parameters from fit to Planck 2015 data. Given are mean values and their 68\% (95\%) confidence
intervals. The right column serves as a comparison to the standard case without lepton asymmetry. The neutrino 
chemical potential $\xi$, dimensionless baryon density $\omega_{\text{b}}$, and the scalar spectral index 
$n_{\text{s}}$ are primary parameters, while the helium mass fraction $Y_{\text{p}}$ and the deuterium abundance 
$y_{\text{DP}}$ are derived parameters.}
\label{means_CL}
\end{largetable}

A large lepton asymmetry can impact the evolution of the Universe at various epochs, e.g. at WIMP decoupling 
\cite{Stuke:2011wz}, the QCD phase transition \cite{Schwarz:2009ii}, big bang nucleosynthesis (BBN) 
\cite{Steigman:2005uz} and photon decoupling \cite{Lesgourgues:1999wu}. In this work, we focus on the last 
two of them, i.e. on the impact of lepton asymmetry on BBN and the CMB. The following arguments hold for 
cosmic temperatures $T$ well below the Pion and Muon masses.

Due to the charge neutrality of the Universe, the asymmetry of \textit{charged} leptons is negligible, because it must be 
of the same order of magnitude as the baryon asymmetry, 
$(n_{\text{e}^{-}}-n_{\text{e}^+})/n_{\gamma}= \mathcal{O}(\eta_{\text{b}})$. 
Neglecting the contribution of charged leptons, today's lepton asymmetry therefore is the sum of the 
neutrino flavour asymmetries $\eta_{\nu,\alpha}$ and can be expressed by the (dimensionless) chemical 
potentials\footnote{The $\xi_\alpha$ are also called degeneracy parameters. This is a misleading name, 
as they are constants only under certain circumstances.} 
$\xi_{\alpha} \equiv \mu_{\alpha}/T_{\nu}$ of neutrinos with flavour $\alpha$,
\begin{equation}
\eta_{\text{l}} \approx \!\!\! \sum_{\alpha=e,\mu,\tau} \!\!\! \eta_{\nu,\alpha} = \frac{1}{12 \zeta(3)} \sum_{\alpha=e,\mu,\tau} \!\!
\left( \frac{T_{\nu,\alpha}}{T_{\gamma}} \right)^3 \!\! \left( \pi^2 \xi_{\alpha} + \xi_{\alpha}^3 \right),
\label{leptonasymmetry}
\end{equation}
where $\zeta(3) \approx 1.20206$. We assumed in equation \eqref{leptonasymmetry} that neutrinos are produced in thermal and chemical equilibrium and that their equilibrium distribution is preserved afterwards. By equation \eqref{leptonasymmetry}, lepton asymmetry is a time-dependent quantity, i.e. its value before electron-positron annihilation ($(T_{\nu,\alpha}/T_{\gamma})^3=1$) differs from its value after electron-positron annihilation by a factor of $(T_{\nu,\alpha}/T_{\gamma})^3=4/11$. This convention for the definition of lepton asymmetry is somewhat misleading, as lepton number is a conserved quantity. A more intuitive definition would include a division by entropy density instead of photon number density, see e.g. \cite{Schwarz:2009ii}. We however use the more conventional definition \eqref{leptonasymmetry} within this work and note that we always refer to values of the lepton asymmetry \eqref{leptonasymmetry} \textit{today}, i.e. after electron-positron annihilation. 

Lepton asymmetry changes BBN in two ways: A chemical potential of the electron neutrino changes the neutron-to-proton ratio at the onset of BBN \cite{Beaudet:1976}, i.e. a positive chemical potential reduces the neutron-to-proton ratio, as an excess of neutrinos can stimulate neutron decay, whereas a negative sign leads to an enhancement at the onset of BBN. Furthermore, lepton asymmetry changes the expansion rate of the Universe. For (effectively) massless neutrinos this can be expressed as an excess of the number of relativistic degrees of freedom,
\begin{equation}
\Delta N_{\text{eff}} = \frac{15}{7} \sum_{\alpha} \left( \frac{\xi_{\alpha}}{\pi} \right)^2 \left[ 2+ \left( \frac{\xi_{\alpha}}{\pi} \right)^2 \right] \geq 0.
\label{Neff}
\end{equation} 
An increased expansion rate results in an earlier freeze-out of weak processes and therefore a higher neutron-to-proton ratio.

The first effect depends only on the \textit{electron} neutrino chemical potential and its \textit{sign}, the second effect 
depends on the \textit{absolute values of all} three chemical potentials. Both effects have an impact on the prediction of the abundances of primordial elements. It turns out that the abundance of primordial helium is a good leptometer, while 
the deuterium abundance is a much better baryometer \cite{Steigman:2005uz}.
 
The CMB is influenced in the presence of lepton asymmetry in the following way. Firstly, an increased $N_{\text{eff}}$ delays the time of matter-radiation equality which leads to an enhanced early integrated Sachs-Wolfe effect. 
Furthermore, a modified value of the amount of helium -- expressed by the helium mass fraction $Y_{\text{p}}$ -- 
alters the tail of the CMB angular power spectrum by the amount of diffusion damping. Lastly, for massive neutrinos 
the neutrino Boltzmann hierarchy is altered in the presence of neutrino chemical potentials \cite{Lesgourgues:1999wu}. In 
this work, we however assume for simplicity that all neutrinos are ultra-relativistic at the time scales of interest, 
such that the Boltzmann hierarchy remains the same as for standard neutrinos\footnote{Note that the Planck collaboration 
assumes in contrast one massive neutrino with $m_{\nu}=0.06$ eV \cite{Ade:2015xua}.}.

\begin{widetext}
\includegraphics[width=\textwidth]{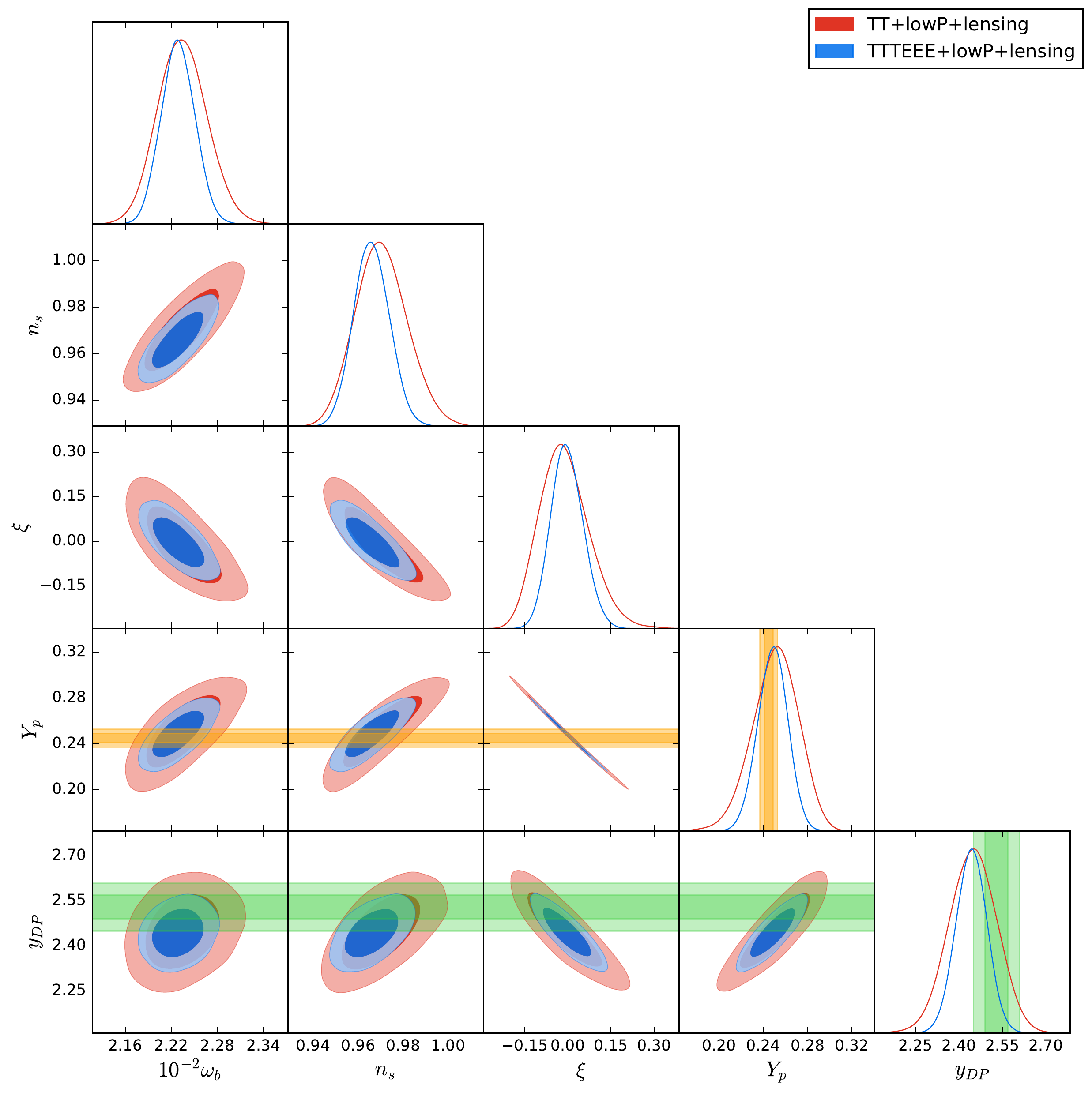}
\caption{2d and 1d marginalised posterior distributions of the cosmological parameters most affected by lepton asymmetry, where the contours are the 68\% and 95\% confidence regions. Orange  and green bands are the 68\% and 95\% confidence regions of direct measurements of the primordial helium (orange, $0.2449 \pm 0.0040$, 68\%, \cite{Aver:2015iza}) and deuterium abundance (green, $y_{\text{DP}}=2.53 \pm 0.04$, 68\%, \cite{Cooke:2013cba}).}
\label{posteriors}
\end{widetext}

In principle, the individual neutrino flavour asymmetries $\eta_{\nu,\alpha}$ and accordingly the corresponding
chemical potentials can have different values. It was however shown in \cite{Dolgov:2002ab,Wong:2002fa,Mangano:2011ip}
that flavour asymmetries are likely to equilibrate before the onset of BBN due to neutrino oscillations, such that finally 
$\xi_{\text{e}} \approx \xi_{\mu} \approx \xi_{\tau}$. Note that this is not generically true, but depends on the neutrino 
mixing angles as well as on the initial values of the neutrino chemical potentials \cite{Johns:2016enc,Barenboim:2016shh}. Particularly, the relation between $\eta_{\text{l}}$ \eqref{leptonasymmetry} and $\Delta N_{\text{eff}}$ \eqref{Neff} does not necessarily hold any longer due to the combined effect of oscillations and collisions around the epoch of neutrino decoupling \cite{Pastor:2008ti,Castorina:2012md}. Within this 
work, we however focus on the case of identical neutrino flavour asymmetries. We furthermore assume that the 
dimensionless chemical potentials are constant at all times of interest, i.e. after neutrino decoupling until the formation of the CMB. 
%This implies that we neglect neutrino masses.

\section{Model fitting and comparison}

\begin{figure}
\centering
\onefigure[width=0.49\textwidth]{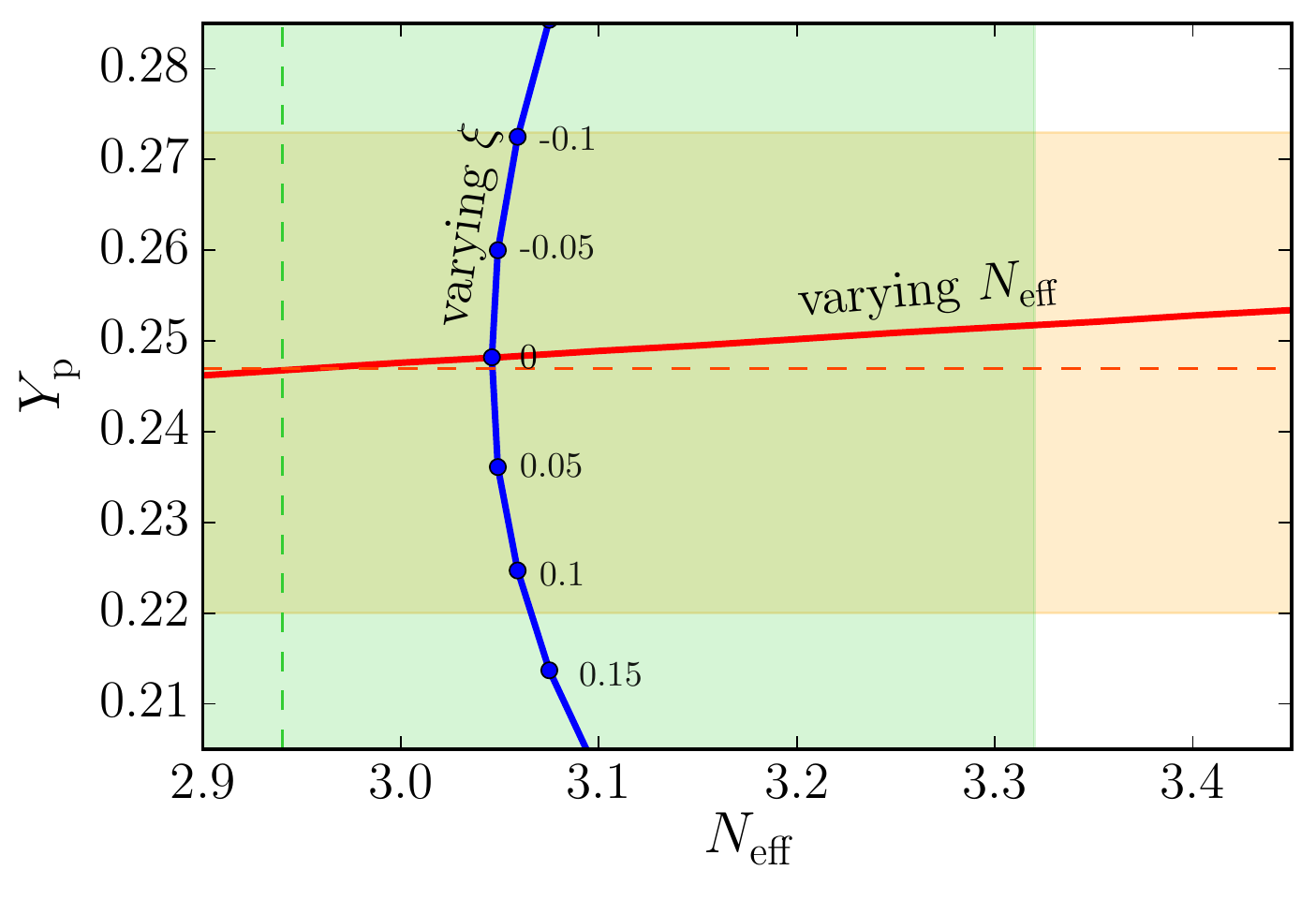}
\caption{BBN prediction for the helium abundance $Y_{\text{p}}$ with $N_{\text{eff}}$ as a free parameter (red solid) and $\xi$ as a free parameter (blue solid), where the blue circles mark different values of $\xi$. Dashed lines are the mean values for $Y_{\text{p}}$ (orange) and $N_{\text{eff}}$ (green) as 1-parameter extensions to the base $\Lambda$CDM model from \cite{Ade:2015xua} (TTTEEE+lensing), shaded regions are the corresponding 95\% confidence regions.}
\label{Yp_Neff}
\end{figure}

We produced a table for the helium mass fraction $Y_{\text{p}}$ using the AlterBBN code 
\cite{Arbey:2011nf}\footnote{version 1.4, https://alterbbn.hepforge.org/}, spanning thereby the parameter 
range $\xi=\lbrace -1.0,1.0 \rbrace$ and $\omega_{\text{b}}=\Omega_{\text{b}}h^2= \lbrace 0.01,0.035 \rbrace$. 
From this table we found a fitting formula that reproduces the tabulated $Y_{\text{p}}$ values with 
precision $\lesssim \mathcal{O}(10^{-4})$, i.e. at the level of the theoretical uncertainties. We implemented 
this fitting formula for $Y_{\text{p}}(\omega_{\text{b}},\xi)$ together with the modified 
$\Delta N_{\text{eff}}(\xi)$ from equation \eqref{Neff} in the Boltzmann code \textsc{class} 
\cite{Blas:2011rf}\footnote{version 2.6.0, http://class-code.net/}. Analogously to $Y_{\text{p}}$ we also obtained a 
fitting function for the deuterium abundance, expressed as $y_{\text{DP}}=10^5 n_{\text{D}}/n_{\text{H}}$, and 
added $y_{\text{DP}}$ as an additional derived parameter in our analysis. For the deuterium abundance we used an 
updated version of the AlterBBN code \cite{Riemer-Sorensen:2017vxj} that includes corrections to nuclear reaction 
rates which lead to significant changes in the predicted deuterium abundance ($\mathcal{O}(5\%)$), but negligible 
changes for helium ($\mathcal{O}(0.1\%)$).

Using the Markov Chain Monte Carlo (MCMC) engine Monte Python \cite{Audren:2012wb}, we explore the cosmological parameter space, which consists of the 6 base cosmological parameters plus the neutrino dimensionless chemical potential, $\lbrace\Omega_{b}h^2, \Omega_{\text{c}} h^2, 100 \theta_{\text{MC}}, \tau, \ln(10^{10}A_{\text{s}}), n_{\text{s}}  \rbrace$ and $\xi$. All parameters are assumed to have flat priors. We use the CMB data from the Planck 2015 release \cite{Ade:2015xua}, where we include the high-$\ell$ temperature data (TT), the high-$\ell$ temperature and polarisation data (TTTEEE), the low-$\ell$ combined polarization and temperature data (lowP) as well as the lensing reconstruction (lensing). 
We adopt the Gelman-Rubin convergence criterion $R<0.01$. 

In figure \ref{posteriors}, we present the posterior distributions of those parameters which are most affected in the presence 
of lepton asymmetry. For better comparison we also plot the 1$\sigma$- and 2$\sigma$-regions of direct measurements 
of the primordial helium ($Y_{\text{p}}=0.2449 \pm 0.0040$, \cite{Aver:2015iza}) and deuterium abundance 
($y_{\text{DP}}=2.53 \pm 0.04$, \cite{Cooke:2013cba}). The corresponding mean values as well as the confidence 
regions of the MCMC analysis can be found in table \ref{means_CL}. 

The constraints on the neutrino dimensionless chemical potential can be translated into constraints on the lepton 
asymmetry via equation \eqref{leptonasymmetry}, i.e.\ $-0.14 \leq\eta_{\text{l}} \leq 0.13$ (TT+lowP+lensing) and 
$-0.085 \leq \eta_{\text{l}} \leq 0.084$ (TTTEEE+lowP+lensing) at 95\% CL.
We can furthermore impose an upper limit on the number of effective degrees of freedom \eqref{Neff}, i.e.\ 
$\Delta N_{\text{eff}}<0.042$ (TT+lowP+lensing) and $\Delta N_{\text{eff}}<0.017$ (TTTEEE+lowP+lensing) at 95\% CL. Remarkably, these constraints are stronger by a factor of at least $\sim10$ than those obtained from an analysis without lepton asymmetry, but with $N_{\text{eff}}$ as a free parameter \cite{Ade:2015xua} (comparing to  $N_{\text{eff}}=2.94\pm 0.38$ for TTTEEE+lensing, 95\% CL). This reveals the importance of taking the full effect of lepton asymmetry on BBN into account -- which includes not only the effect of $\Delta N_{\text{eff}}$ but also the modified weak interactions that change the neutron-to-proton ratio. Figure \ref{Yp_Neff} illustrates this aspect in more detail: We show the BBN prediction for the helium mass fraction $Y_{\text{p}}$ for the case of $N_{\text{eff}}$ as a free parameter (red) and the case of $\xi$ as a free parameter (blue). The orange shaded region presents the 95\% confidence region of $Y_{\text{p}}$ for an analysis \cite{Ade:2015xua} with $Y_{\text{p}}$ as an \textit{additional free parameter} to the base $\Lambda$CDM parameters. While the orange region constrains the red curve only very weakly, the blue curve is tightly constrained to a region of small values for $N_{\text{eff}}$ (and therefore small $\xi$). See also \cite{Schwarz:2012yw} and \cite{Hamann:2007sb} for a discussion about the importance of BBN consistent analyses.

With an agreement to zero lepton asymmetry within 1$\sigma$ we conclude that there is no evidence for lepton asymmetry from CMB data. Note that this however still leaves a lot of space for a lepton asymmetry orders of magnitude larger than the baryon asymmetry. 

It is furthermore interesting to note that the inclusion of lepton asymmetry enhances the errors of $\omega_{\text{b}}$, 
$Y_{\text{p}}$, $y_{\text{DP}}$ and $n_{\text{s}}$ compared to the minimal 
$\Lambda$CDM model (see the right column in table \ref{means_CL}). Since in the standard model $Y_{\text{p}}$ and $y_{\text{DP}}$ are only functions of $\omega_{\text{b}}$, their errors are usually also entirely determined by the error of $\omega_{\text{b}}$. It is therefore natural that the additional dependence of $Y_{\text{p}}$ and $y_{\text{DP}}$ on $\xi$ results in an enhancement of the uncertainties of $Y_{\text{p}}$, $y_{\text{DP}}$ and $\omega_{\text{b}}$. The enhancement in the error of $n_{\text{s}}$ can be explained by the degeneracy of $Y_{\text{p}}$ with $n_{\text{s}}$. 

Our constraints for $\xi$ are at least a factor of $\sim 4$ stronger than previous constraints in \cite{Caramete:2013bua}\footnote{Note however that the sum of neutrino masses $\sum m_{\nu}$ is additionally allowed to vary in \cite{Caramete:2013bua}.} (using Planck data) and \cite{Shiraishi:2009fu,Popa:2008tb,Schwarz:2012yw} (using WMAP data). Comparing our results to the Planck forecast \cite{Hamann:2007sb} we can confirm the expected 68\% confidence region. 
Our constraints are weaker by a factor of $\sim 2$ than the constraints from \cite{Simha:2008mt,Mangano:2011ip,Cooke:2013cba}, where direct measurements of primordial elements were used \textit{additionally} to the CMB constraints for $\omega_{\text{b}}$. Note however that those constraints from direct measurements should be treated with caution: Direct measurements of the primordial helium abundance seem to suffer from significant systematic errors, comparing e.g. $Y_{\text{p}}=0.2551\pm 0.0022$ \cite{Izotov:2014fga} with the measurement reported above and shown for comparison in the figures of this work, $Y_{\text{p}} = 0.2449 \pm 0.0040$ \cite{Aver:2015iza}. Even though weaker, the CMB constraints derived in this work therefore seem to be more robust.

\begin{figure}
\centering
\onefigure[width=0.45\textwidth]{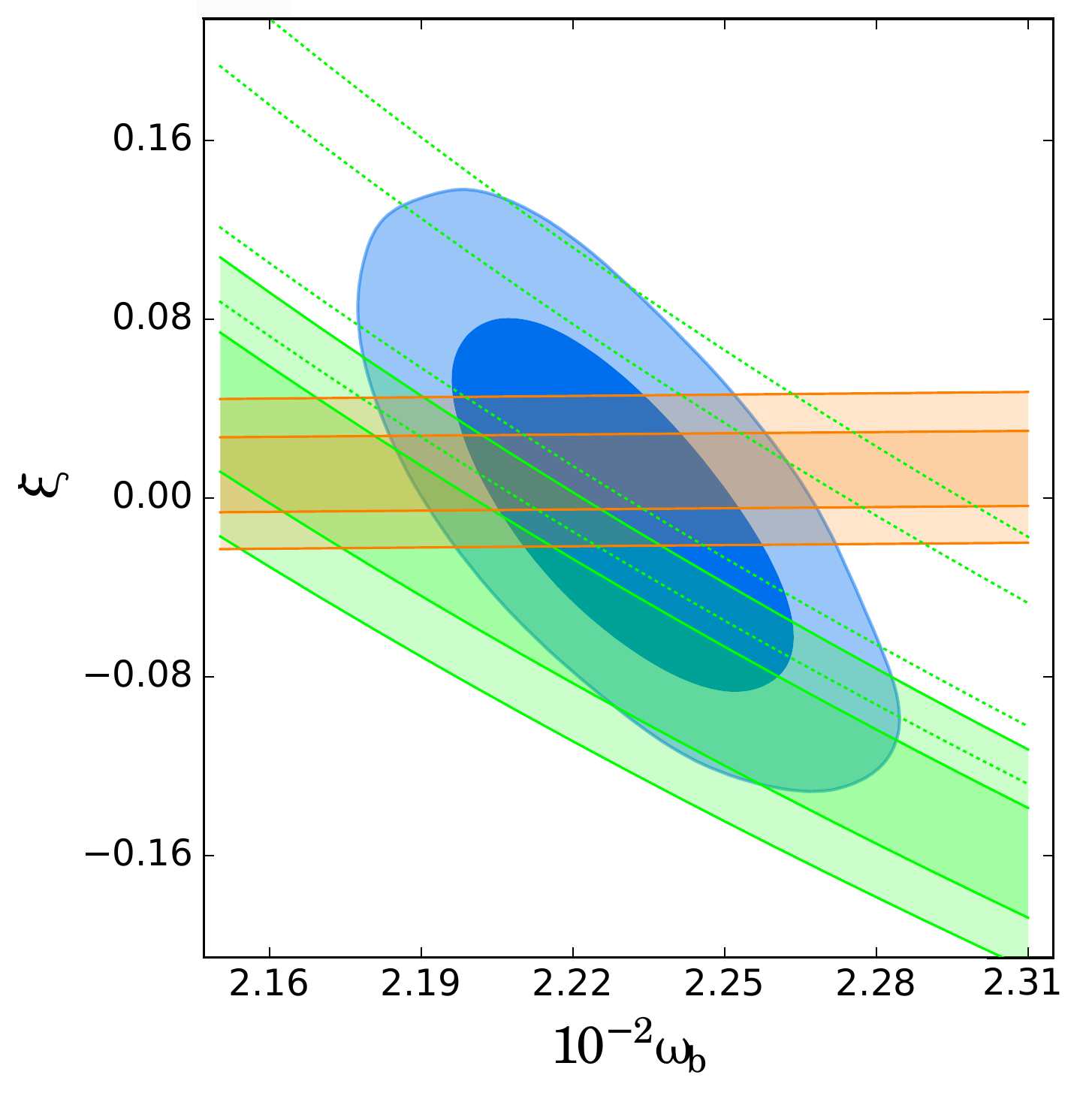}
\caption{2d marginalised posterior distribution in the $(\xi,\omega_{\text{b}})$-plane using Planck 2015 TTTEEE+lowP+lensing data (blue), direct measurements of the primordial helium (orange) \cite{Aver:2015iza} and deuterium (green) \cite{Cooke:2013cba} abundance, where the contours are the 68\% and 95\% confidence regions. Dotted green lines show the constraints from the deuterium measurements without the update of the AlterBBN code \cite{Riemer-Sorensen:2017vxj}.}
\label{xi_omegab}
\end{figure}

Our derived constraints on the primordial helium and deuterium abundance are consistent within 2$\sigma$ 
with those from direct observations, i.e. $Y_{\text{p}} =0.2449 \pm 0.0040$ (68\% CL) \cite{Aver:2015iza} and 
$y_{\text{DP}}=2.53 \pm 0.04$ (68\% CL) \cite{Cooke:2013cba}. The constraints from direct measurements can be 
translated into constraints in the $(\xi,\omega_{\text{b}})$-plane, as we see in figure \ref{xi_omegab}. Being 
almost perpendicular to the $\xi$-axis, the form of the helium contour (orange) shows why helium is generally a 
good \textit{leptometer} \cite{Steigman:2005uz}: The constraints on $\xi$ are relatively independent of constraints 
on the baryon density. Interestingly, the overlap of the CMB data contours in figure \ref{xi_omegab} (blue) with the 
deuterium measurement (green) reveals a tendency towards negative values of the neutrino chemical potential, i.e.\ 
towards a negative lepton asymmetry. This tendency is however not observed, when the previous version 
the of AlterBBN code 
is used (green dotted). This manifests how important the correct treatment of nuclear reaction rates is and is another indication that the systematic uncertainties from BBN exceed the statistical uncertainties.

A rigorous comparison of our constraints to direct measurements of $y_{\text{DP}}$ and $Y_{\text{p}}$ would  however demand a more careful treatment of the systematic uncertainties of the BBN predicted values: 
First of all, we did not correct for the small difference between the definition of the helium fraction in BBN codes ($m_{\text{He}} \equiv 4 m_{\text{N}}$) and helium fraction in Boltzmann codes\footnote{The Planck 2015 analysis takes the mass defect into account, whereas the Planck 2013 analysis \cite{Ade:2013zuv} also neglects it.}, which account for the 
mass defect of helium (a $\mathcal{O}(1\%)$ correction). Furthermore, we did not include the theoretical uncertainties of $Y_{\text{p}}$ ($\mathcal{O}(0.05 \%)$) and $y_{\text{DP}}$ ($\mathcal{O}(5\%)$) due to uncertainties in the measurements 
of nuclear reaction rates. 
Finally, a more rigorous approach would also take into account the fact that there exist discrepant measurements of the neutron life time, $\tau_{\text{n}}=\left[880.3\pm1.1\right] \mathrm{s}$ \cite{Agashe:2014kda} and $\tau_{\text{n}}=\left[887.8\pm 1.2 (\text{stat.}) \pm 1.9 (\text{syst.}) \right] \mathrm{s}$  (68\% CL) \cite{Yue:2013qrc}, that lead to an additional theoretical uncertainty in $Y_{\text{p}}$ ($\mathcal{O}(0.6\%)$) and $y_{\text{DP}}$ ($\mathcal{O}(0.4\%)$)\footnote{Our fitting formula for $Y_{\text{p}}$ and $y_{\text{DP}}$ are based on calculations with the AlterBBN default value \mbox{$\tau_{\text{n}}=885.7$ s.}}. For non-zero chemical potentials this uncertainty can increase to maximally $\mathcal{O}(1\%)$, i.e. the relation between $Y_{\text{p}}$ and $\xi$ predicted by BBN is correct at the level of $\mathcal{O}(1\%)$. Note that current CMB data with a precision of $\mathcal{O}(5-10\%)$ 
in $Y_{\text{p}}$ (see table \ref{means_CL}) are not sensitive to such small differences in $Y_{\text{p}}$, and all of our results in table \ref{means_CL} should be mainly unaffected by those additional uncertainties. For a comparison with direct measurements of $Y_{\text{p}}$ and $y_{\text{DP}}$ (that have errors of only $\mathcal{O}(2 \%)$) those additional uncertainties become however relevant and should be taken into account. 

\section{Conclusions}
In this work, we derived constraints on lepton asymmetry using the Planck 2015 data \cite{Ade:2015xua} of the CMB angular power spectra. 
We modified the Boltzmann code \textsc{class} in order to take into account the increased number of relativistic degrees of freedom \eqref{Neff} and the modified helium fraction $Y_{\text{p}}(\omega_{\text{b}},\xi)$ to impose BBN consistency. We thereby assumed equal flavour asymmetries and neglected neutrino masses. We find $\xi=-0.011 ^{+0.179}_{-0.167}$ (95\% CL) using the high-$\ell$ temperature data, the low-$\ell$ polarization data and lensing reconstruction (TT+lowP+lensing). Including the high-$\ell$ polarization data (TTTEEE+lowP+lensing) reduces the uncertainty of $\xi$ further, resulting in $\xi=-0.002 ^{+0.114}_{-0.111}$ (95\% CL). For the lepton asymmetry $\eta_{\text{l}}$ \eqref{leptonasymmetry} this implies $-0.14 \leq\eta_{\text{l}} \leq 0.13$ (TT+lowP+lensing) and $-0.085 \leq \eta_{\text{l}} \leq 0.084$ (TTTEEE+lowP+lensing) at 95\% CL. Our constraints are significantly stronger than previous constraints from CMB data \cite{Popa:2008tb,Shiraishi:2009fu,Schwarz:2012yw,Caramete:2013bua}. Even though our CMB constraints are weaker (factor $\sim 2$) than the constraints obtained from direct measurements of light element abundances \cite{Simha:2008mt,Mangano:2011ip,Cooke:2013cba}, we think that they are more robust, since direct measurements of primordial light elements are 
still subject to sizeable systematic uncertainties.
Our derived constraints on the primordial helium $Y_{\text{p}}$ and deuterium $y_{\text{DP}}$ abundance are in good agreement with direct measurements \cite{Aver:2015iza,Cooke:2013cba}. The 2d marginalised posteriors in 
the $(\omega_{\text{b}},\xi)$-plane \mbox{(fig. \ref{xi_omegab})} might hint at some tension between the CMB and the observed light 
element abundances, but systematic issues like BBN nuclear reaction rates become relevant at that level of precision.

Finding consistent constraints on lepton asymmetry at BBN and CMB epochs provides by itself a non-trivial consistency 
check of the hot big bang model. On the other hand, $\eta_{\text{l}}$ is still allowed to exceed $\eta_{\text{b}}$ by many orders of magnitude, which could have interesting consequences for physics at $T > T_{\text{BBN}}$  \cite{Schwarz:2009ii,Stuke:2011wz,Barrie:2017mmr}.

\acknowledgments
We thank T.~Tram for his support with the Monte Python software package, E.~Jenssen for providing the updated AlterBBN code, and D.~Boriero, L. Johns, S. Pastor and M.~Wygas for useful discussions and comments. IMO acknowledges the support by Studienstiftung des Deutschen Volkes. We acknowledge support from RTG 1620 “Models of Gravity” funded by DFG. Numerical calculations have been performed using the computing resources from RWTH Compute Cluster.

\bibliographystyle{eplbib}
\bibliography{Literature}

\end{document}